\documentclass{emulateapj}

\input psfig.sty

\slugcomment{ApJ, in press}

\shorttitle{The MRI in stirred disks}
\shortauthors{Workman \& Armitage}

\begin{document}

\title{Interaction of the magnetorotational instability with \\ hydrodynamic turbulence 
in accretion disks}

\author{Jared C. Workman\altaffilmark{1,2} and Philip J. Armitage\altaffilmark{1,2}}
\altaffiltext{1}{JILA, Campus Box 440, University of Colorado, Boulder CO 80309; 
workmanj@colorado.edu, pja@jilau1.colorado.edu}
\altaffiltext{2}{Department of Astrophysical and Planetary Sciences, University of Colorado, Boulder CO 80309}

\begin{abstract}
Accretion disks in which angular momentum transport is dominated by the 
magnetorotational instability (MRI) can also possess additional, purely 
hydrodynamic, drivers of turbulence. Even when the hydrodynamic processes, 
on their own, generate negligible levels of transport, they may still 
affect the evolution of the disk via their influence on the MRI. 
Here, we study the interaction between the MRI and hydrodynamic turbulence 
using local MRI simulations that include hydrodynamic forcing. As expected, 
we find that hydrodynamic forcing is generally negligible if it yields a 
saturated kinetic energy density that is small compared to the value 
generated by the MRI. For stronger hydrodynamic forcing levels, we find that 
hydrodynamic turbulence modifies transport, with the effect varying depending upon 
the spatial scale of hydrodynamic driving. Large scale forcing boosts transport by an 
amount that is approximately linear in the forcing strength, and leaves the character of the 
MRI (for example the ratio between Maxwell and Reynolds stresses) unchanged, 
up to the point at which the forced turbulence is an order of magnitude 
stronger than that generated by the MRI. Low amplitude small scale forcing may 
modestly suppress the MRI. We conclude that the impact 
of hydrodynamic turbulence on the MRI is generically ignorable in cases, 
such as convection, where the additional turbulence arises due to the 
accretion energy liberated by the MRI itself. Hydrodynamic turbulence may affect 
(and either enhance or suppress) the MRI if it is both strong, and driven by independent 
mechanisms such as self-gravity, supernovae, or solid-gas interactions in 
multiphase protoplanetary disks.  
\end{abstract}

\keywords{accretion, accretion disks --- hydrodynamics --- instabilities --- MHD --- turbulence}

\section{Introduction}
The magnetorotational (MRI) or Balbus-Hawley instability \citep{vel59,chandra61,balbus91,balbus98} underpins  
the most important --- and possibly only --- source of outward angular momentum 
transport in a wide class of well-ionized accretion disks. The MRI destabilizes disk flows 
in which ${\rm d}\Omega^2 / {\rm d}r < 0$, and leads to a state of sustained 
magnetohydrodynamic (MHD) turbulence that transports angular momentum outward 
\citep{hawley95,brandenburg95,armitage98,hawley00,papaloizou03,hirose06}. The 
majority of the transport is mediated by Maxwell rather than Reynolds stresses. 
Following convention, the efficiency of angular momentum transport within 
disks is measured via an equivalent Shakura-Sunyaev (1973) 
$\alpha$ parameter, which can be expressed in terms of the fluctuating 
velocity and magnetic fields as,
\begin{equation}
\alpha P_0 = \rho v_r \delta v_\phi + \frac{B_r B_\phi}{\mu_0},
\label{alpha}
\end{equation}
where $P_0$ is the thermal pressure.
The first term on the right-hand side of this equation represents the Reynolds 
(or fluid) stress and the second term represents the Maxwell (or magnetic) stress. 
There is no strict equivalence between the evolution of MHD turbulent disks 
and models that assume an $\alpha$ shear viscosity \citep{balbus98,pessah07}, 
but for our purposes $\alpha$ is a convenient measure of the efficiency 
of angular momentum transport.

The fact that the MRI dominates the transport of angular momentum within 
accretion flows does not, of course, imply that other sources of turbulence 
do not exist within disks. The most striking example occurs in galactic 
disks, in which turbulence can be driven by thermal instabilities and 
supernova explosions in regions that are unstable to the MRI \citep{piontek05}. 
However, even in ``normal" accretion disks around stars or compact objects 
there are numerous possibilities, including self-gravity in sufficiently 
massive disks \citep{toomre64}, convection \citep{stone96}, and Kelvin-Helmholtz 
instabilities excited by the interaction between gaseous and solid components of 
protoplanetary disks \citep{cuzzi93}. Additional fluid motions can also be driven within 
disks due to the gravitational influence of embedded planets \citep{bate02} or 
binary companions \citep{spruit87}, though the wave-like fluid motions 
induced by these sources are different from those initiated by turbulence. Some of 
these processes may drive fluid motions within the disk whose kinetic energy 
density is comparable to that resulting from the MRI. In particular, a self-gravitating 
disk can remain stable against fragmentation while generating an effective 
$\alpha \simeq 10^{-1}$ \citep{gammie01,rice05} --- as large 
or larger than that produced by the MRI. Since the fluid motions resulting 
from hydrodynamic processes are uncorrelated with the MRI, it is reasonable 
to suspect that there could be non-trivial interactions between the MRI 
and other turbulent processes in disks within which both are operating 
simultaneously. Indeed, \cite{fromang04} found that in a self-gravitating, 
MRI-turbulent disk, the strength of the angular momentum transport from 
the self-gravity was both weaker, and had a different time-dependence, when  
compared to a disk in which self-gravity alone was at work. This result 
is striking, since the disk simulated by \cite{fromang04} was dominated 
by low-order (azimuthal wavenumber $m=2$) self-gravitating structure whose scale 
was much larger than the most unstable MRI scales.

In detail, different drivers of hydrodynamic turbulence may influence the 
MRI in a unique manner, requiring a case-by-case study (self-gravity, for example, 
is a special case since it {\em does} yield outward transport of angular momentum, 
whereas most other hydrodynamic mechanisms yield negligible or even inward transport). 
We do not address such subtleties here, but rather study how generic driven hydrodynamic 
turbulence of specified strength interacts with the MRI. One's expectation 
is obviously that hydrodynamic turbulence that is weak (say, in terms of 
the saturated value of the kinetic energy density) compared to that driven 
by the MRI ought to leave the MRI unscathed, with significant interaction 
developing when the two sources of turbulence are of comparable strength. 
Very strong turbulence can amplify magnetic fields in the 
disk {\em independent} of the MRI, though this may not necessarily be accompanied 
by angular momentum transport. Our goal in this paper is to test such 
order of magnitude intuition.

The focus of this paper is on small-scales, which can be captured most 
effectively using local shearing-box simulations. Such simulations have 
significant limitations that must be borne in mind. In particular, the 
strength of angular momentum transport in zero-net flux simulations with 
purely numerical viscosity and magnetic diffusivity has a marked 
dependence on numerical resolution \citep{gardiner05,fromang07a}. Moreover, 
when physical values for the transport coefficients {\em are} used (as in this paper) the 
strength of turbulence depends upon the magnetic Prandtl number 
$Pm \equiv \nu / \eta$ as well as on the Reynolds number \citep{lesur07,fromang07b}. 
What this means for real disks --- which when highly ionized have 
values of the diffusivity $\eta$ and viscosity $\nu$ that are much 
smaller than can currently be simulated --- is unclear, but an obvious 
implication for numerical experiments is that the absolute value 
of $\alpha$ derived from shearing-box simulations must be viewed with 
caution. For the time being, constraints
derived from modeling of observed systems may be
more reliable \citep{kpr07}. 
For this reason, our focus here is on how the strength of 
angular momentum transport varies in the presence of additional 
hydrodynamic turbulence, rather than its absolute value.

The plan of this paper is as follows. In \S2 we describe 
the set-up for our runs, which make use of the {\sc PENCIL} MHD code 
previously employed for both disk and turbulence calculations \citep{haugen04}. 
In \S3 we present results, which concentrate on the influence of 
different levels of hydrodynamic turbulence on the saturation level, 
structure, and angular momentum transport efficiency of the MRI. These 
results are summarized and discussed in \S4.

\section{Methods} 
We use the {\sc Pencil} Code\footnote[1]{The {\sc Pencil} Code and its documentation are available at
http://www.nordita.org/software/pencil-code/} to solve the dynamical equations used to describe a local, magnetized
patch of an accretion disk in the co-rotating frame.  The {\sc Pencil} Code solves the equations of compressible MHD in
non-conservation form using sixth order spatial derivatives and third order temporal derivatives. The {\sc Pencil} Code
has been used previously to study the MRI \citep{brandenburg95}, and has been shown to yield similar 
results to calculations performed using {\sc ZEUS} and similar algorithms \citep{hawley95,balbus98}. One 
important difference between {\sc Pencil} and {\sc ZEUS} is that {\sc Pencil} requires the use of explicit diffusive terms to maintain stability. 
{\sc Pencil} is a fixed grid, Eulerian code, that is parallelized using MPI.

\subsection{Equations} 
We follow the motion of a magnetized parcel of isothermal gas in the co-rotating frame. A local, 
Cartesian co-ordinate system is used, in which $x$ represents the radial direction, $y$ the 
azimuthal direction, and $z$ the vertical direction. The
equations of motion describing the system are derived by solving for deviations in the flow velocity $\mathbf{u}$ from the
Keplerian shear flow $U_y(x)=(-3/2)\Omega x$, where $\Omega$ is the local background rotation rate
\citep{brandenburg95,wisdom88}.  

Using the framework described above, the momentum equation is given by

\begin{eqnarray}
\frac{\partial \mathbf{u}}{\partial t} &=&  -u_y^{(0)}\frac{\partial \mathbf{u}}{\partial y}-(\mathbf{u
\cdot \nabla u})+2\Omega u_y - \frac{1}{2}\Omega u_x \nonumber \\
&-&\frac{ \mathbf{\nabla}P}{\rho} + \frac{\mathbf{J \times B}}{\rho} + \vec{F} + \nu_h\nabla^6\mathbf{u}
\label{momentum}
\end{eqnarray}

\noindent

\noindent 

where $\mathbf{u}$ is the velocity vector, $P$ is the pressure, $\rho$ is the density, $\mathbf{J}$ is the current,
$\mathbf{B}$ is the magnetic field, and $u_y^{(0)}$ is the background Keplerian shear. The terms in equation
(\ref{momentum}) correspond to: advection due to the background flow,  advection of the perturbed velocity 
field,  Coriolis
and shear effects, pressure gradients, the Lorentz force, our forcing term (to be described in the next section) and
viscous terms.  We neglect the effects of vertical gravity both for the sake of computational economy, 
and because it is likely irrelevant for this study focusing on the interplay between
the MRI and forced hydrodynamic turbulence. The viscous term $\nu_h\nabla^6\mathbf{u}$ describes 
so-called hyperviscous dissipation, specified by a parameter $\nu_h$. 
We use hyperviscosity to limit the effects of dissipation to the smallest
scales and thereby extend our inertial range. For reviews of the effects of hyperviscosity on MHD
simulations and a derivation of the hyperviscous term see \citet{brandenburg2002} and \citet{johansen2005}.

The continuity equation is, 

\begin{equation}
\frac{\partial \rho}{\partial t} =   -u_y^{(0)}\frac{\partial\rho}{\partial y} - \mathbf{\nabla \cdot}(\rho\mathbf{u})
\label{mass}
\end{equation}

\noindent 
where the first term comes from advection due to the background flow and the second term refers to the standard mass flux.   We use upwinding rather than diffusive terms to stabilize the mass conservation equation.

The final equation describes the evolution of the magnetic vector potential. {\sc Pencil} 
solves for the vector potential
$\mathbf{A}$ rather than the magnetic field $\mathbf{B} = \mathbf{\nabla \times A}$, which ensures
a solenoidal (divergence free) magnetic field. The equation for the evolution of the vector potential is given by 

\begin{equation}
\frac{\partial \mathbf{A}}{\partial t} =   -u_y^{(0)}\frac{\partial\mathbf{A}}{\partial y}  + \frac{3}{2}A_y\mathbf{\hat x} + \mathbf{u  \times (\nabla \times A)} + \eta_h \nabla^6\mathbf{A},
\label{vectorpot}
\end{equation}
where the first term describes advection of the potential due to the background flow, the second term describes
magnetic stretching  due to shear, the third term describes the standard electromotive force, and the last term is the hyperdiffusive resistivity.  

The final equation needed to close the system is the energy equation. We use an isothermal equation of state
so this is simply $P=\rho c_s^2$, where $c_s$ is the sound speed.

\subsection{Forcing Term}

Our goal in applying hydrodynamic forcing is to
mimic the effects of hydrodynamic turbulence that is generated within the 
disk by mechanisms that are independent of the MRI.
There are many possible approaches to accomplishing
this, that vary depending upon whether the forcing is
applied throughout the volume or only at the boundaries,
and in the temporal and spatial scales involved. The ``best" 
approach evidently depends upon the physics one seeks to 
mimic -- forcing that is gravitationally driven, for example, 
would logically be written as the gradient of a potential function. 
For most of our runs we 
use a forcing term that injects energy throughout the
volume at large scales (physically, this would be
roughly the disk scale height). In the hydrodynamic
regime this energy then cascades down to smaller
scales where it is eventually dissipated (by viscosity at very small physical 
scales in a real disk, by hyperviscous dissipation at the grid scale in the 
simulations. We have also considered the effect of forcing the fluid 
on small spatial scales. In either case we adopt the forcing 
function introduced in Haugen et. al. (2004) which is given by

\begin{equation}
F(\mathbf{x},t) = \mathbf{f_k}N e^{[i\mathbf{k}(t)\mathbf{\cdot x} + i \phi (t)]}.
\label{force}
\end{equation} 
In this function the wavevector $\mathbf{k}(t)$ and $\phi (t) \epsilon [-\pi,\pi]$ are chosen randomly at every time
step guaranteeing that the forcing due to the function is uncorrelated in time.  In order to ensure that the
energy injection rate is independent of the time step is it normalized (using dimensional arguments) by setting,
\begin{equation} 
N=f_0c_s^{3/2}\sqrt{|\mathbf{k}|/\delta t},
\end{equation}
where the scalar $f_0$ defines the amplitude of the 
resulting forcing. We discuss how we set $f_0$ in the next Section. 
At each time step a random wavevector $\mathbf{k}$ is selected from a
certain range in $k$ space along with a unit vector $\mathbf{e}$. These vectors are then used 
in the construction of the argument of the exponent as well as in the creation of 

\begin{equation}
\mathbf{f_k}=\frac{\mathbf{k \times e}}{\sqrt{k^2-\mathbf{(k\cdot e)^2}}}
\end{equation}
which describes a nonhelical, transverse wave.

\subsection{Setup}

We choose a box sized $(2\pi)^3$ with a resolution of $160^3$. The boundary conditions in the y and z directions are
periodic and those in the x direction are shearing. Shearing sheet boundary conditions are described in 
\citet{hawley95}.  We initialize the system with a density of $\rho=1$, a sound speed $c_s=1$, and an angular
rotation rate of $\Omega = 0.2$.  This scaling returns a box size of $2 \pi H/5$.  We include hyperviscosity to stabilize the scheme at the grid scale. The
coefficients for the fluid viscosity and the magnetic resistivity 
are set equal to each other yielding a (hyperviscous / hyperresistive) Prandtl number ($\nu_h/\eta_h$) of 1 for most
simulations. Our hyperviscous and hyperresistive Reynolds numbers for grid scale shocks are 75.  
One technical point to be borne in mind when comparing our results with other MRI simulations is 
that we use a cubical domain rather than one elongated in the azimuthal direction. A cubical domain 
maximizes the range between the dissipation scale and the forcing scale, which necessarily has power 
on the {\em smallest} dimension of the domain. The uniform resolution in the three spatial 
directions ensures that hyperviscous and hyperresistive dissipation act isotropically. We do 
not see any significant differences between our results and prior simulations that we could 
ascribe to the differing simulation domains.

For our first simulation we initiate the MRI by setting up a magnetic field in the z direction that has the form $A_0
\cos (kx)\mathbf{\hat z}$ where $A_0= 0.05$. This yields a plasma $\beta$ (defined as the ratio of the gas pressure
to the magnetic pressure) of $\beta=800$ at the peak of the magnetic field. The instability is seeded by 
perturbing the velocity field with random gaussian fluctuations whose magnitude is of the order of 
$10^{-3} c_s$. With this setup the most unstable wavelength for the MRI, given by $\lambda_c = 2\pi V_{a_z} / \sqrt{3} \Omega$ 
\citep{balbus91}, is resolved across 23 grid points. 
We do not use any forcing for the initial run which is intended to allow the MRI to develop into a saturated state.
After evolving the simulation to 20 orbits we stop it at this point and create a restart dump. From here we restart
the MRI only simulation and let it run for $\sim$130 more orbits. We call this simulation $0f_0$ and it is our
'fiducial' run with which we compare all subsequent runs.

\subsection{Large Scale Forcing}
Once the MRI simulation is completed we begin several sets of hydrodynamic simulations with equivalent parameters but
no magnetic fields.  To these simulations we add the forcing function given by equation (\ref{force}).  We force the
simulations around  $k = 1.5$ to simulate the effects of generic hydrodynamic turbulence being input at scales slightly smaller than our simulation domain.  
We force slightly interior to the boundaries to avoid issues with the boundary conditions.  
The forcing injects energy into the box at large scales, which then cascades down in the normal hydrodynamic 
manner.  We
run hydrodynamic simulations with varying values for $f_0$ until the time averaged level of kinetic energy in the
hydrodynamic simulations, due to the forcing function, approximately matches the time averaged level of kinetic energy in our 
MRI only run $0f_0$. We label our hydrodynamic only run as HL and allow it to run for  $\sim$150 orbits.  

Having determining the appropriate value for $f_0$ we initialize six runs starting from the fully saturated MRI state we
saved at 20 orbits in the MRI only case. We implement forcing from this point onward in each of the simulations with values of
$[\frac{1}{2}f_0,1f_0,2f_0,4f_0,8f_0,16f_0]$ and run the simulations for approximately 130 orbits each.  We label each run by the value of
$f_0$; HL is the purely hydrodynamical (large scale) run, $0f_0$ is the fiducial MRI only run, $.5f_0$ corresponds to the run with
half strength forcing, and so on. The details of each run are summarized in Table~\ref{modelparameters}.

\begin{table*}
\center
\begin{tabular}{llllllllll}
\hline
{Model} & {MHD} & {$f_0$} & {$\beta$$^{(a)}$} & {Grid} & {Size} &  
{$\lambda_c/\delta_z$$^{(b)}$} & 
{Orbits} & {$\mathbf{<\alpha >_{\delta t,v}}$$^{(c)}$}
 & {$\nu/\eta$} \cr
\hline
\hline
HL & No & 0 & 800  & $160^3$ & $(2\pi)^3$ & 23.1 & 150 & -1e-7 & NA\cr
$0f_0$ & Yes & 0 & 800  & $160^3$ &$(2\pi)^3$ & 23.1 & 150 & .00132 & 1\cr
$.5f_0$ & Yes & $\frac{1}{2}$ & 800  & $160^3$ &$(2\pi)^3$ & 23.1 & 150 & .00127 &  1\cr
$1f_0$ & Yes & $1$ & 800  & $160^3$ &$(2\pi)^3$ & 23.1 & 150 & .00163 & 1\cr
$2f_0$ & Yes & $2$ & 800  & $160^3$ &$(2\pi)^3$ & 23.1 & 150 & .00187  & 1\cr
$4f_0$ & Yes & $4$ & 800  & $160^3$ &$(2\pi)^3$ & 23.1 & 150 & .00319 & 1\cr
$8f_0$ & Yes & $8$ & 800  & $160^3$ &$(2\pi)^3$ & 23.1 & 150 & .00510 & 1\cr
$16f_0$ & Yes & $16$ & 800  & $160^3$ &$(2\pi)^3$ & 23.1 & 150 & .00764 & 1\cr
$2f_0l$ & Yes & $2$ & 800  & $160^3$ &$(2\pi)^3$ & 23.1 & 150 & .0021 & 1\cr
$8f_0l$ & Yes & $8$ & 800  & $160^3$ &$(2\pi)^3$ & 23.1 & 150 & .0048 & 1\cr
$2f_0p4$ & Yes & $2$ & 800  & $160^3$ &$(2\pi)^3$ & 23.1 & 150 & .0031 & 4\cr
$2f_0p.25$ & Yes & $2$ & 800  & $160^3$ &$(2\pi)^3$ & 23.1 & 150 & .0023 & .25\cr
HS & No & 0 & 800  & $160^3$ & $(2\pi)^3$ & 23.1 & 150 & 4.2e-5 & NA\cr
$0sf_0$ & Yes & 0 & 800  & $160^3$ &$(2\pi)^3$ & 23.1 & 109 & .0016  & 1\cr
$1sf_0$ & Yes & $1$ & 800  & $160^3$ &$(2\pi)^3$ & 23.1 & 124 & .0008  & 1\cr
$2sf_0$ & Yes & $2$ & 800  & $160^3$ &$(2\pi)^3$ & 23.1 & 115 & .00078  & 1\cr
$4sf_0$ & Yes & $4$ & 800  & $160^3$ &$(2\pi)^3$ & 23.1 & 115 & .0026  & 1\cr
\hline
\end{tabular}
\caption{Model Parameters. Notes: $^{(a)}$The maximal value of $ \beta =P_{gas}/P_{magnetic}$, at $t=0$; 
 $^{(b)}$$\lambda_c = 2\pi V_{a_z} / \sqrt{3} \Omega$; $^{(c)}$denotes a time and volume average from the 20th orbit until the simulation's
conclusion}
\label{modelparameters}
\end{table*}

In addition to the simulations described above we also ran four others with some of the underlying physics modified. 
The goal of these runs was to test how robust our results are to plausible changes in the forcing function 
(which is essentially arbitrary) and plasma microphysics. 
In two simulations we changed the time scale over which the
forcing function switches to new wave vectors.  The function is still defined and normalized as described by Equation
(\ref{force}) except that new vectors are now selected at intervals of $P =
2\pi/\Omega$. This is the natural resonance timescale for the disk. 
We run two simulations with this altered forcing, both starting at the fiducial 20 orbital periods and
running out until 150 periods, one having $2f_0$ and the other $8f_0$ as their forcing amplitudes.  We label these runs
$2f_0l$ and $8f_0l$, where the $l$ indicates the long term nature of the forcing function.

Our final two large scale forcing simulations are motivated by the work of \citet{lesur07} and \citet{fromang07b} 
who demonstrate a link
between the value of the magnetic Prandtl number $\nu/\eta$ and the saturation level of magnetic turbulence.  We initialize
two forced runs at 20 orbits with a $2f_0$ forcing amplitude (as this seems to correspond to the demarcation between the
standard MRI behavior and that influenced by the hydrodynamic forcing) 
and run them to 150 orbits.  For one run, labeled $2f_0p4$ we increase the
hyperviscous coefficient by a factor of four to correspond to a four fold increase in the Prandtl number.  For the other
run, labeled $2f_0p.25$ we reduce the value of the  hyperresistive coefficient to correspond to a four fold decrease in the
Prandtl number.

\subsection{Small Scale Forcing}
In addition to examining the effects of large scale forcing 
we explore a (smaller) range of simulations where we input energy at a much smaller scale. 
Once again we set up a non forced MRI simulations 
with conditions equivalent to those described above. 
We run this to 20 orbits and save the values at this point. 
We then allow the unforced MRI to run to $\sim 110$  orbits and label this run $0sf_0$ where the $s$ indicates 
the small scale forcing.

We then repeat the process, described above, of determining the value of the forcing strength parameter $sf_0$
which produces a saturated kinetic energy level in a purely
hydrodynamical run which is approximately equal to that produced in the fiducial MRI run, except now 
forcing at $k=40$.  This value is chosen as it represents small scales within our inertial range which are not yet affected
by dissipation due to numerical effects. Having obtained an approximate match between the kinetic energy 
densities in the MRI-only and hydrodynamic runs, we label the forcing amplitude as $sf_0$ and implement a series of forced runs starting from the fiducial 20
orbit point using multiples of this value. These runs are labeled $1sf_0,2sf_0,$ and $4sf_0$.

\section{Results}
Due to the fluctuating nature of turbulent quantities both spatial and temporal averaging 
is needed in order to obtain reproducible values for quantities such as the saturated 
magnetic field strength \citep{winters03,papaloizou03}. We present results in terms 
of the volume average as a function of time, given by
\begin{equation}
<F(t)> = \frac{1}{V}\int F(t) dV,
\label{volumeavg}
\end{equation} 
and in terms of a running time average given by 
\begin{equation}
<F>_{\delta t} = \frac{1}{t-t_0} \int _{t_0}^t < F(t)> dt. 
\label{timeavg}
\end{equation}
We use the latter quantity to assess whether our runs are long enough to attain 
convergence. We note that our run length has been chosen to be just 
sufficient to yield clear discrimination between the different forcing cases --- much 
longer runs would be required to measure quantities to percent level precision.

\subsection{Large Scale Forcing Runs}

\subsubsection{Fiducial Run $(0f_0)$}
\begin{figure}
  \plotone{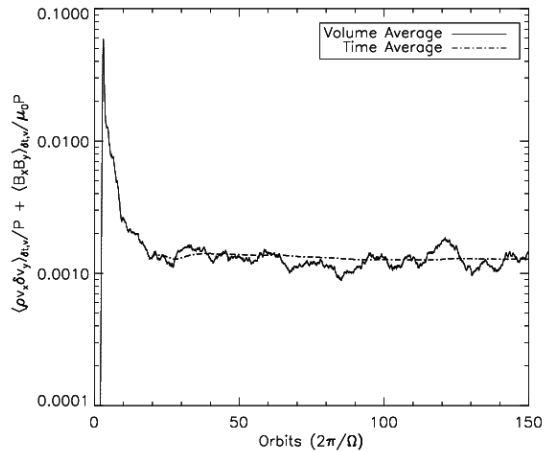}
  \caption{Time and volume averages (equations~\ref{timeavg} and~\ref{volumeavg}) of the transport
coefficient, $\alpha$ (equation~\ref{alpha}) for the fiducial MRI run ($0f_0$).}
  \label{fig:mrialpha}
\end{figure}

We compare all of our runs to a fiduical non-forced MRI configuration. This run displays the characteristic behavior of
MRI simulations \citep{brandenburg95, balbus98}.  There is an initial exponential growth phase which amplifies all
components of the magnetic and velocity fields, which reach peak amplitude at $\sim 3$ orbits. After the peak there is a
gradual reduction in all quantities until they saturate at a (fluctuating) level several orders of magnitude above their
initial levels (except for $B_Z^2$ which, in these zero-net flux simulations, saturates at a level 
below its initial value). The saturated state is sustained and all quantities remain approximately constant 
until we stop the run at orbit $\sim150$.  

Figure~\ref{fig:mrialpha} shows the total transport coefficient $\alpha$ for the MRI only case. The 
behavior is as described above: an exponential increase in $\alpha$ followed by a
decay to reach a saturated state. The saturated value of $\alpha$ is around $1.3 \times 10^{-3}$. 
As has been found in all prior simulations, the level of transport associated
with the magnetic stresses is almost an order of magnitude higher than that associated with the fluid 
stresses. 

The MRI-only run serves two purposes. First, we use it as a reference for comparison with the 
forced runs. For this comparison, we compute running averages from the MRI-only run starting 
at orbit 20, thereby excluding the initial (unphysical) exponential growth and decay phase. 
Second, we use the snapshot generated at orbit 20 as an initial condition for the forced 
runs. These forced runs thereby model the effect of additional turbulence on an already 
saturated disk. We do not study how strong forcing would impact the linear growth phase of the 
MRI, as it seems unlikely that this is relevant to any physical system.

\subsubsection{Hydro Run $(HL)$}
\begin{figure}
  \plotone{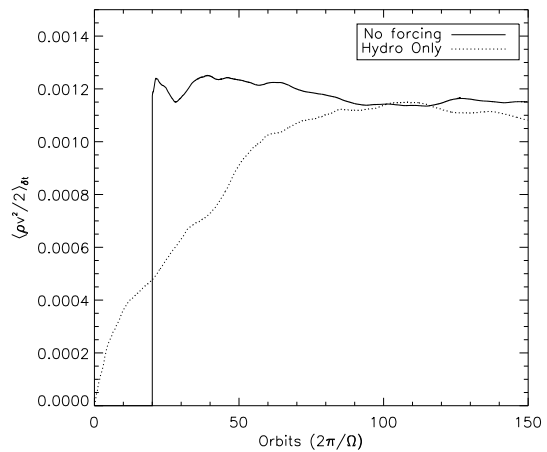}
  \caption{Running time averages (equation~\ref{timeavg}) of the kinetic 
  energy for the unforced MRI run ($0f_0$) and the hydrodynamic run (HL). 
  In the hydrodynamic run forcing was used. The time averages for the MRI 
  run are begun at 20 orbits to prevent contamination from the initial exponential 
  phase of the instability. The mean value of the time averages agree to within $\sim 10\%$ after orbit 80.}
  \label{fig:hydromri}
\end{figure}

We next ran several forced,  non-MHD simulations, in order to find the value of the forcing 
amplitude (defined through $f_0$) that yields hydrodynamic turbulence of the same strength as that 
obtained in the MRI-only run. MHD and non-MHD turbulence can be physically different \citep{sridhar94}, 
so there is no unique definition of equivalent strength. We define an equivalent hydrodynamic 
run as one which yields close to the same value of the saturated kinetic energy density as the MRI-only 
run\footnote{Alternatively, we could define equivalence via the rate of energy {\em input}. This, 
however, is harder to measure in the MRI case.}. Numerically, we found that a value of $f_0 = 0.0035$ 
yielded reasonable agreement (at a level of about 10\% between orbits 80 and 150). For this matched 
case Figure~\ref{fig:hydromri} shows how the running 
time averages of the kinetic energy compare in the MRI-only and hydro-only runs. Convergence of the 
kinetic energy in the box is rather slow in the hydrodynamic case (possibly due to the low level 
of viscosity, since lower resolution test runs converged more rapidly), but the level of the 
agreement at the end of the runs is sufficient for our purposes.

\begin{figure}
  \plotone{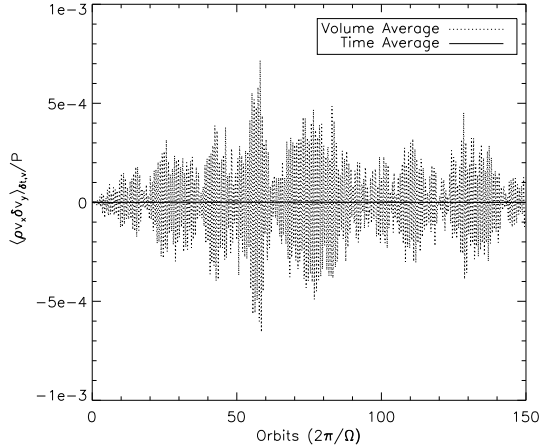}
  \caption{Time and volume averages (equations~\ref{timeavg} and~\ref{volumeavg}) 
  of the transport coefficient, $\alpha$ (equation~\ref{alpha}) for the purely 
  hydrodynamical, large scale forcing simulation (HL). No transport is induced 
  by the hydrodynamic forcing when it is employed on its own.}
  \label{fig:hydroalpha}
\end{figure}

Figure~\ref{fig:hydroalpha} shows the running time average of the fluid $\alpha$ in the 
hydrodynamic only reference run with this value of $f_0$. As expected for randomly forced 
hydrodynamic turbulence, the level of transport obtained is consistent with zero and 
certainly negligible compared with either the Maxwell or Reynolds stresses generated 
in the MRI-case. This result is important as we are explicitly looking at the effects of large
scale hydrodynamic turbulence on the MRI. If a particular forcing function were to generate a non-zero value for
$\alpha$ it would be difficult to separate its effects on transport from that of the MRI.

\subsubsection{Forced Runs $(\frac{1}{2}f_0,1f_0,2f_0,4f_0,8f_0,16f_0)$}

Having determined the level for $f_0$ that matches the kinetic energy between the 
MRI-only and hydro-only runs, we initiate six MHD runs with the amplitude of the forcing 
set at $\frac{1}{2}f_0$, $1f_0$, $2f_0$, $4f_0$, $8f_0$, and $16f_0$. We are particularly interested in 
the behavior of the magnetic fields and the angular momentum transport obtained in these simulations. 

\begin{figure} 
\plotone{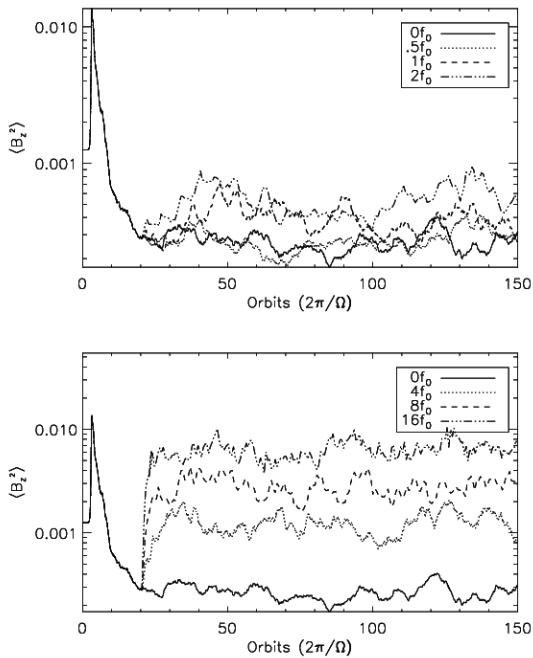} 
\caption{Evolution of $B_z^2$, averaged over the simulation volume,  
  for all MHD simulations including large scale forcing ($0f_0, 1f_0, 2f_0, 4f_0, 8f_0, \& \, 16f_0$). 
  All forcing strengths above the $1 f_0$ case induce noticeable increases in field strength.}
 \label{fig:bzbz}
\end{figure}

Figure~\ref{fig:bzbz} shows the time evolution of the volume average of $B_z^2$ 
for the fiducial unforced MRI simulation, together with the results from the 
six simulations with variable forcing levels. The amplitude of the vertical magnetic 
field component is statistically similar (given the large fluctuations) for the 
$\frac{1}{2}f_0$ and $1f_0$ cases as compared to the unforced MRI-only run. 
For stronger forcing ($2f_0$ and above) $B_z^2$ clearly exceeds that which would 
be generated by the action of the MRI alone. Very roughly the energy in the vertical 
field scales linearly with the amplitude of forcing (recall that the forcing amplitude is defined 
via the energy in fluid motions).

Our interpretation of the behavior of $B_z^2$ in the presence of forcing is that 
the existence of external hydrodynamic turbulence introduces more turnover in the vertical
direction via random fluid motions. This increase in the ``kneading" of the $z$ 
component of the magnetic field by the forced turbulence increases $B_z^2$. Since the 
generation of vertical field by the MRI is relatively weak (this is reflected in the 
fact that, at saturation, only a small fraction of the magnetic energy in MRI 
simulations is in the $z$ component) it is no surprise that modest levels of 
isotropic hydrodynamic turbulence suffice to generate higher $B_z^2$.

\begin{figure}
  \plotone{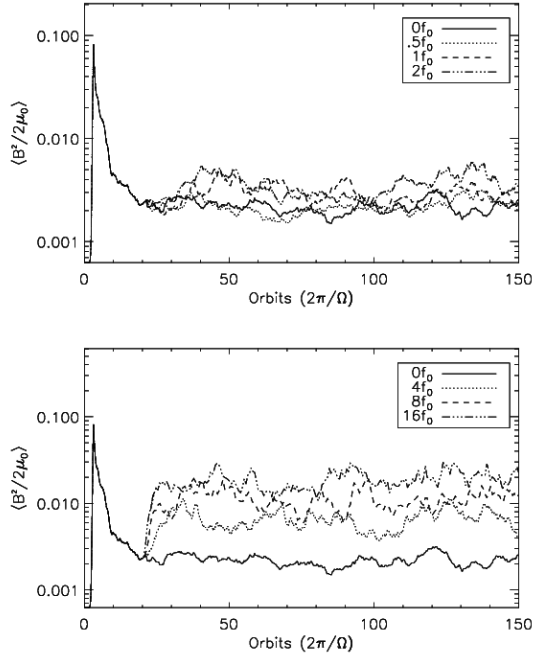}
  \caption{Evolution of $B^2/2\mu_0$, the total energy in the magnetic field within the simulation volume. 
Results are shown for the unforced and all forced MHD runs in the large scale forcing simulations ($0f_0, 1f_0,
2f_0, 4f_0, 8f_0, \& \, 16f_0$). For forcing amplitudes greater than $4f_0$ we see a sharp rise in the strength of the
magnetic field consistent with the idea that a new MRI exponential growth phase is initiated, subsequent to 
which the new
saturation level is maintained.}
  \label{fig:bsq}
\end{figure}

Figure \ref{fig:bsq} displays the total magnetic field energy density for all the MHD simulations.  For all
runs we see an increase in the total field energy as $f_0$ increases. For the weaker 
forcing simulations (those up to $2 f_0$) the timescale over which the disk adjusts 
to a new quasi-steady state (with a higher saturation level) is relatively long -- 
typically tens of orbital periods. Somewhat different behavior is seen in the very 
strongly forced runs ($4 f_0$ and above). In these cases the field adjusts almost 
immediately to a higher value once the forcing is turned on, and these stronger fields 
are then sustained on average for the duration of the simulation. What may be happening 
here is that the enhanced $B_z$ created by the forced turbulence has ``resparked" the 
rapid growth phase of the MRI allowing for more rapid adjustment of the field configuration.

The presence of stronger magnetic fields in the stirred runs raises the question 
of whether the resulting turbulence (i) retains the character of the MRI, (ii) is more 
similar to non-magnetic turbulence, or (iii) is distinct from either. We have 
attempted to quantify the character of the turbulence in two ways. First, we 
study the global characteristics of the turbulence, concentrating on the strength 
of angular momentum transport and how that stress is partitioned between Maxwell 
and Reynolds components. A notable feature of the MRI is that the stress is 
dominated by the Maxwell contribution, although the Reynolds stress is also 
non-zero. Second, we look at the structure of the turbulent flow as measured 
by the power-spectrum of various quantities.

\begin{figure}
  \plotone{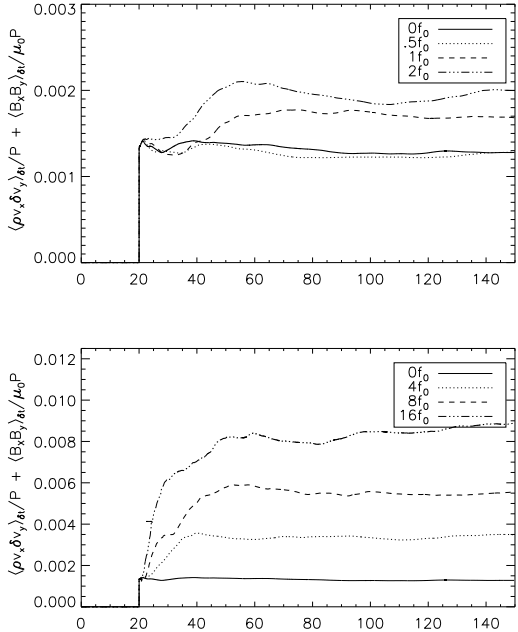}
  \caption{Running time average of $\alpha$ for all large scale forcing, MHD simulations
  ($0f_0,1f_0,2f_0,4f_0,8f_0, \& \, 16f_0$).}
  \label{fig:alphatime}
\end{figure}

\begin{figure}
  \plotone{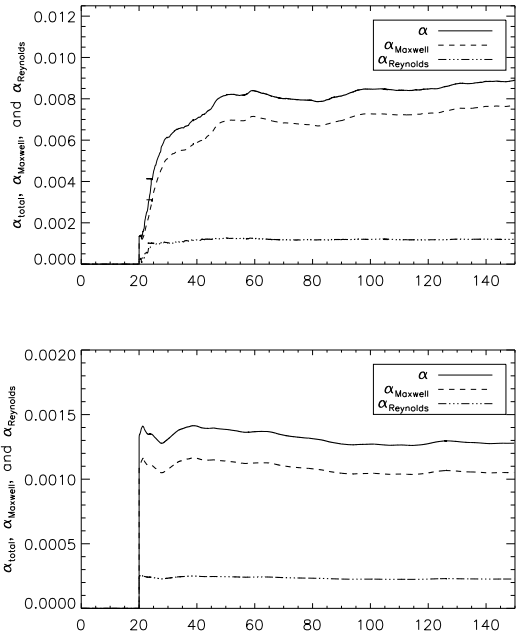}  
  \caption{Running time average of the Maxwell and Reynolds components of $\alpha$
   as well as the total $\alpha$ for the unforced MRI simulation and the run with 
   the strongest large scale forcing ($16 f_0$). 
   The magnetic component of the transport remains dominant in both the
   unforced MRI run and in the most strongly forced case.}
  \label{fig:alpha}
\end{figure}

Figure~\ref{fig:alphatime} shows the time evolution of the equivalent $\alpha$ 
parameter for all of the runs that incorporate large-scale forcing. The division 
between magnetic and hydrodynamic stress is shown in Figure~\ref{fig:alpha} for the 
unforced ($0 f_0$) and most strongly forced ($16f_0$) cases. Forcing boosts the 
strength of angular momentum transport once it is as strong ($1 f_0$) or stronger 
than the MRI in terms of the saturated equivalent kinetic energy. The derived 
$\alpha$ increases steadily with the forcing level up to the $16f_0$ run, 
which displays a value of $\alpha$ about 6 times larger than the unforced MRI-only 
simulation. The fraction of the transport contributed by Maxwell stress as compared to 
Reynolds stress remains surprisingly constant 
between the unforced and maximally forced cases. Even though we are imposing external 
forcing -- in some cases at quite a vigorous level -- the transport is still determined by the magnetic
field, and appears to retain the characteristic behavior expected from the MRI.

\begin{figure} 
  \plotone{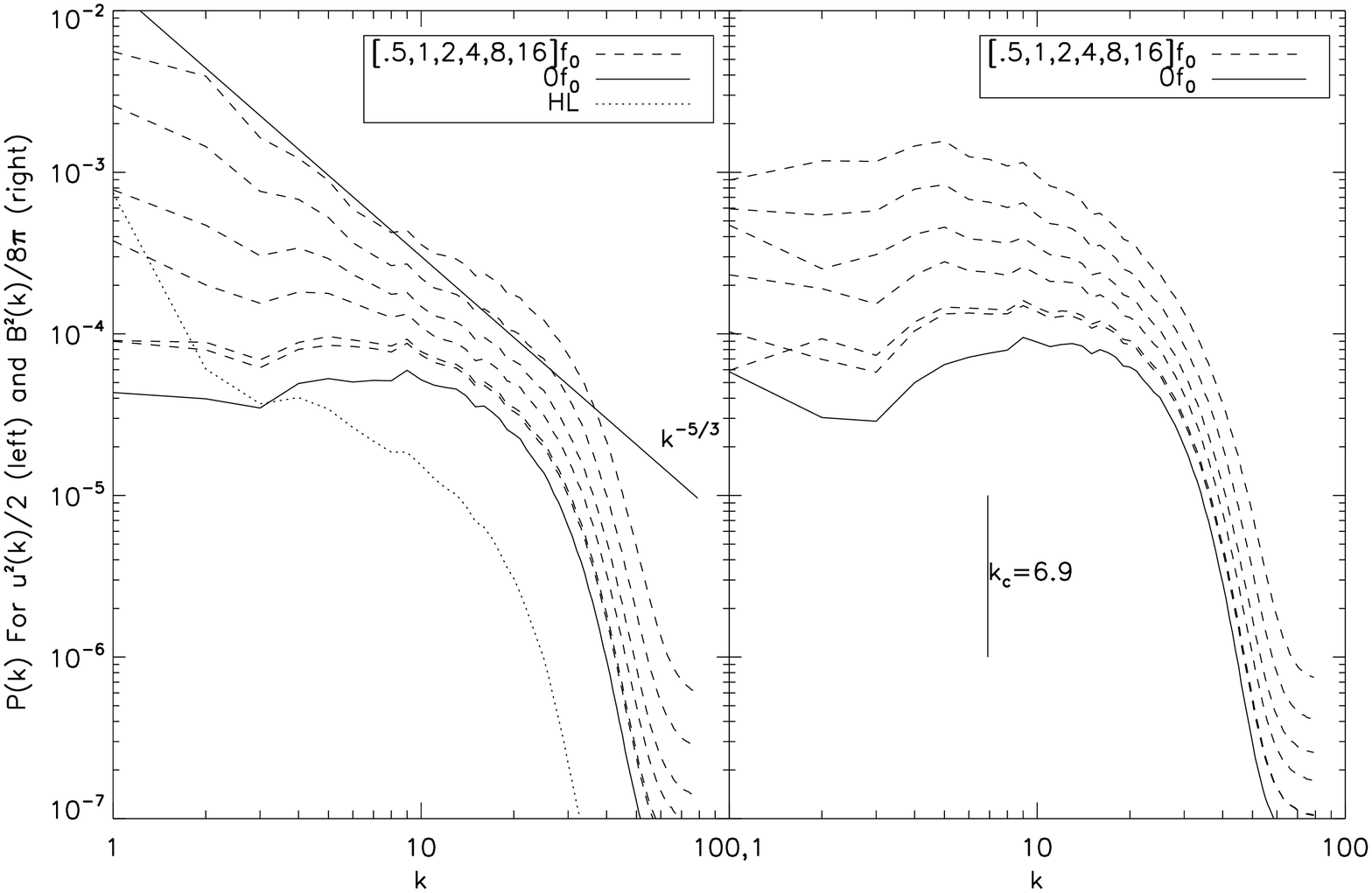}
  \caption{Power spectra of the kinetic energy (left panel) and the magnetic field energy density (right panel) in
the large scale forcing simulations. The spectra have been averaged over orbits 130-140. HL is the hydrodynamic
only large scale forcing simulation, $0f_0$ is the unforced  MRI simulation and the remaining curves show the
forced, large scale simulations. The expected Kolmogorov slope has been plotted for the kinetic energy spectra and
the location of the critical wavelength has been plotted in the panel corresponding to the magnetic field
spectra.  In all cases the total energy in the system increases with forcing amplitude. There is a 
transition between MRI-like and hydrodynamic behavior visible in the kinetic energy spectra at large 
spatial scales.}
  \label{fig:fftl}
\end{figure}

We can also quantify the structure of the turbulence via power spectra of the velocity and 
magnetic fields. Figure~\ref{fig:fftl} shows power spectra for both the kinetic energy and 
(where appropriate) magnetic energy density for the hydrodynamic reference simulation, for 
the unforced MRI run, and for all of the large-scale forcing simulations. For the hydrodynamic 
simulation (shown in the left panel Figure~\ref{fig:fftl}), we see that the kinetic energy 
is dominated by the wavelength which corresponds roughly to the box size at which power is 
injected. For wavenumbers between $k=3$ and $k=30$ the spectrum is approximately of 
the Kolmogorov form ($\propto k^{-5/3}$). For the MRI-only run the spectra of the kinetic 
energy and magnetic fields are quite similar -- they are essentially flat at large scales 
($k < 20$) with a subsequent sharp decline toward the dissipation scale. When forcing 
is included in the MRI simulation we see the same pattern of behavior discussed above 
in the context of $\alpha$. For the relatively weakly forced runs ($0.5 f_0$ and $1 f_0$) 
both the kinetic and magnetic power spectra are only slightly modified from the MRI-only 
case. For stronger forcing we find that the large-scale {\em hydrodynamic} structure of the 
turbulence (quantified by the kinetic energy power spectrum) increasingly resembles that 
of non-magnetized turbulence. For the $16f_0$ run the large-scale slope is reasonably 
close to the Kolmogorov expectation. The power spectrum of the magnetic fields, on the 
other hand, looks like a scaled version of the MRI-only run even up to the strongest 
forcing levels. There is some evidence for a shift of the peak power toward lower 
$k$, which would be consistent with the MRI operating at larger scales in the 
presence of stronger magnetic fields. We interpret these results as suggesting that 
the {\em magnetic} character of the turbulence remains similar (at all scales) to 
that generated by the MRI even in the presence of strong hydrodynamic forcing. The 
stronger magnetic fields generated by adding forcing then merely scale up the 
angular momentum transport efficiency from the value expected in the absence of 
such effects.

 \begin{figure*}
  \includegraphics[scale=.75]{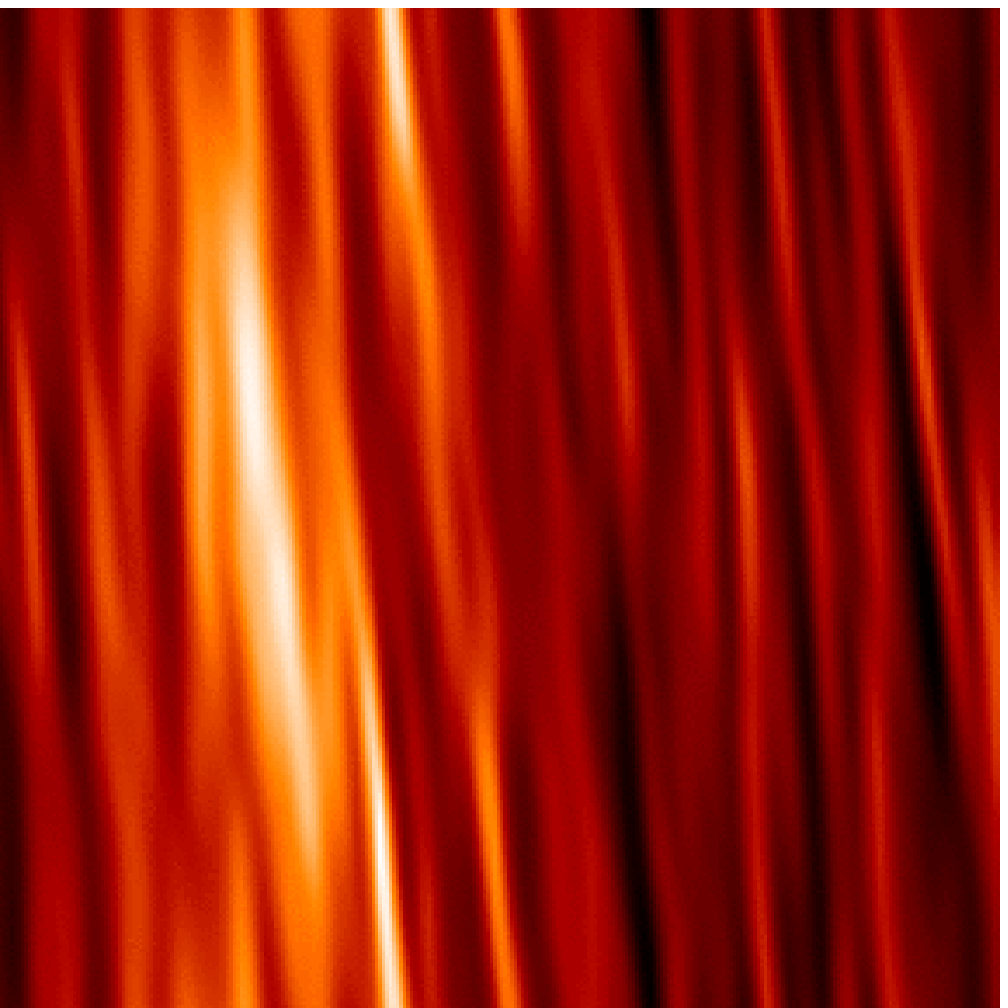}%
  \includegraphics[scale=.75]{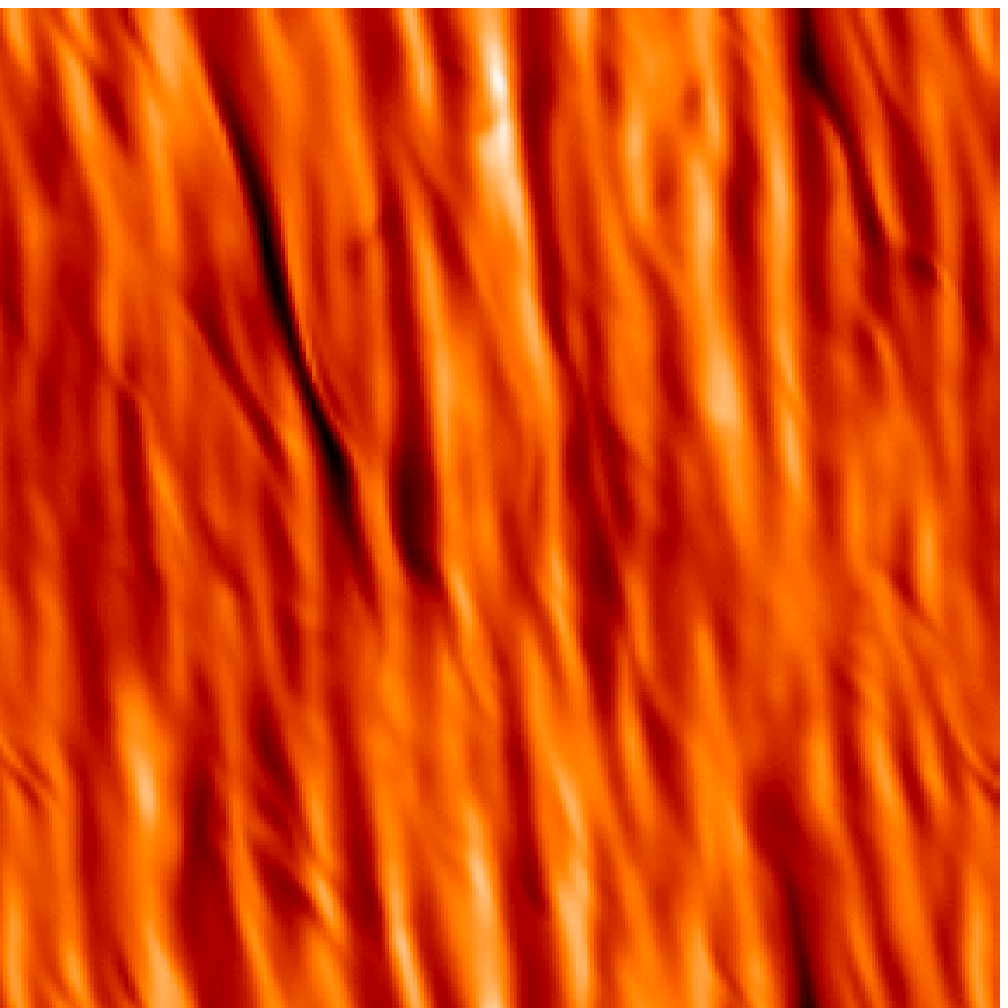}
   \includegraphics[scale=.75]{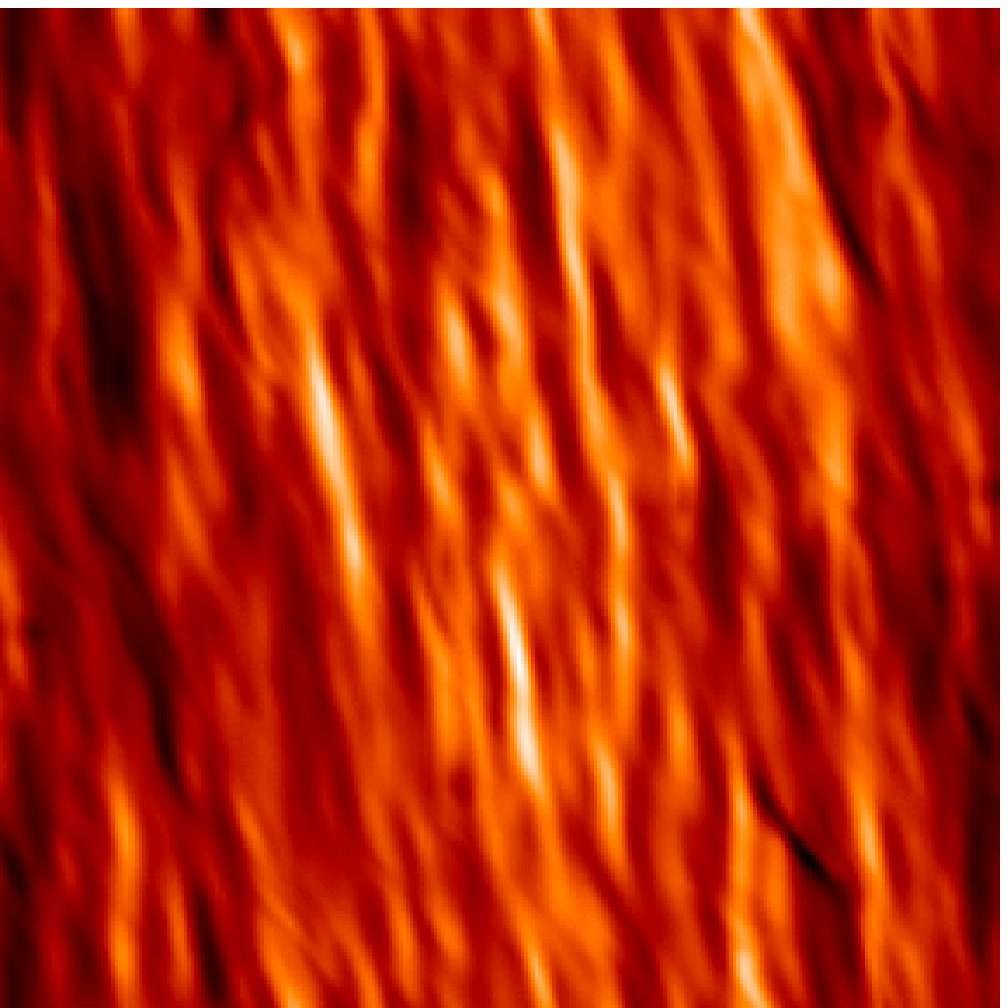}%
  \includegraphics[scale=.75]{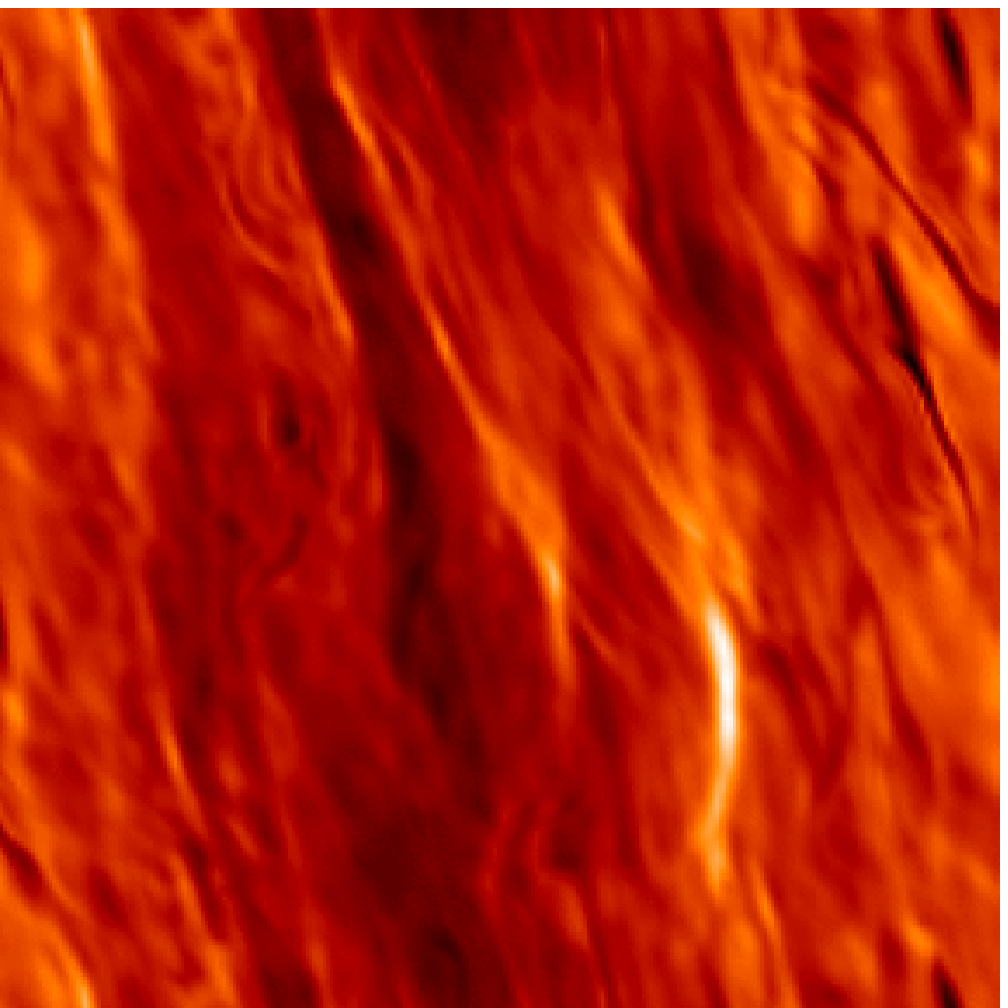}
  \caption{Density slices at 150 orbits in the $xy$ plane taken from simulations 
  HL (upper left), $0f_0$, $4f_0$, \& $16f_0$.}
 \label{rhos}
 \end{figure*}

Figure~\ref{rhos} shows a slice of the density profile in the x-y (r-$\phi$) plane at 
150 orbits for simulations HL, $0f_0$, $4f_0$,and $16f_0$. The dominant effect of the 
shear is clearly visible -- the turbulence is stretched out into ribbons elongated in the 
azimuthal direction. Comparing the plots, about the only feature that is visually  
apparent is the increase in the large-scale power (here in the density field) as the strength 
of hydrodynamic forcing is increased. The most strongly forced run displays substantially 
greater density contrasts, and it is clear that the large-scale morphology of the flow 
has been altered significantly from the MRI-only case. Nevertheless, as discussed above, 
on the (somewhat smaller) scales that are most important for angular momentum transport 
the MHD properties of the flow are apparently not significantly modified.

\begin{figure}
  \plotone{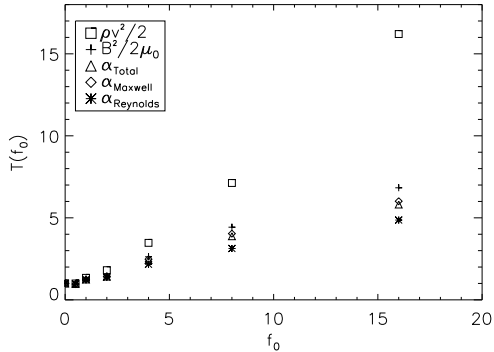}
  \caption{The average values of the magnetic field energy density, kinetic energy and components 
  of $\alpha$ from the large scale forcing runs, normalized to the results of the unforced 
  MRI simulation. The forcing strength is represented on the $x$-axis by the value of the 
  normalization parameter for the forcing $f_0$.}
  \label{fig:tf0}
\end{figure}

To quantify the relationship between the strength of imposed forcing and the resulting physical quantities 
($\alpha$, kinetic energy density etc) we measure the ratio of the mean quantities in the forced to the 
unforced run. This ratio, defined as,
\begin{equation} 
 T(f_0) = \frac{<F(t_{\rm final}, f_0)>_{\delta t}}{<F(t_{\rm final}, f_0=0)>_{\delta t}}
\end{equation} 
is plotted in Figure~\ref{fig:tf0} for $\alpha$, its decomposition into Maxwell and Reynolds 
stresses, the total magnetic field energy, and the total kinetic energy. 
It is evident that, to a good approximation, the additional energy injected via the 
forcing simply raises the saturated value of the kinetic energy in the same manner 
as one would expect for purely hydrodynamic runs -- i.e. the $16 f_0$ run,  
which would have 16 times the kinetic energy of the MRI-only run in the hydrodynamic limit, 
does indeed yield that enhancement in the presence of MHD. The fact that these are MHD runs, 
in which the dissipation properties of turbulence could in principle differ 
significantly from the hydro only case, does not appear to make a substantial 
difference. This is similar to the situation studied in the context of molecular 
clouds --- where MHD and hydrodynamic turbulence have comparable decay rates  
\citep{macclow98,stone98} --- though here the fluid motions are highly subsonic.

As noted previously, all of the magnetic quantities, together with the Reynolds 
stress (which is an approximately fixed fraction of the total stress) increase in 
unison as the forcing strength increases. The relationship is roughly linear in the 
value of the forcing parameter $f_0$, though with a smaller slope than the 1:1 
relationship seen for the kinetic energy. A forcing strength that yields an order 
of magnitude increase in the saturated kinetic energy density results in about a 
factor 4 increase in the the magnetic field energy density and in all of the 
components of $\alpha$.

\subsubsection{Large Scale Runs With Different Prandtl numbers}
A number of recent studies provide evidence that the magnetic Prandtl number, 
\begin{equation}
 {\rm Pm} \equiv \frac{\nu}{\eta}
\end{equation}
is an important parameter whose value affects the saturation level of MHD 
turbulence initiated by the MRI \citep{lesur07,fromang07b}\footnote{The value 
of ${\rm Pm}$ also affects the properties of non-rotating, driven MHD 
turbulence \citep{schekochihin04,iskakov07}.}. This dependence is of interest since 
the predicted value of ${\rm Pm}$ varies substantially between the inner 
regions of black hole accretion disks (where ${\rm Pm} \gg 1$) and the 
outer region \citep{balbus08}. In all of the simulations performed above 
our {\em hyperviscous} Prandtl number $\nu_h/\eta_h$ was equal to one.
To explore the affects of varying the the Prandtl number we ran two simulations, one in which
we increased the value of our viscosity coefficient by a factor of four and one in which we reduced
our resistive coefficient by a factor of four. This change in coefficients allowed for an effective
span of 16 in Prandtl number. We ran both of these simulations with a forcing amplitude of $2f_0$ as
this marked the lowest forcing level for which significant differences with the unforced 
runs were observed. We ran these simulations from our fiducial start point at 20 orbits out to 150
orbits and labeled them $2f_0p4$ and $2f_0p.25$.

\begin{figure}
  \plotone{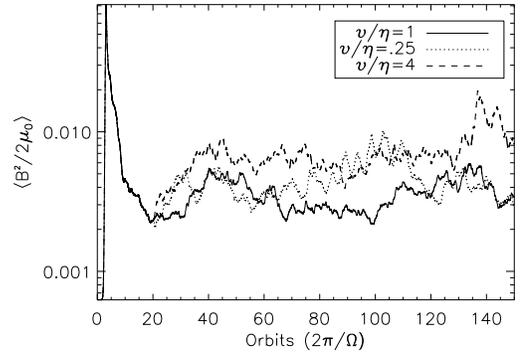}
  \caption{Evolution of the magnetic field energy density ($B^2/2\mu_0$) for the
simulations with altered Prandtl numbers ($\nu/\eta$). All three runs used $2f_0$ as the forcing amplitude.}
  \label{fig:magvolp}
\end{figure}

Figure~\ref{fig:magvolp} shows the volume averaged magnetic energy density for simulations $2f_0$,
$2f_0p4$, and $2f_0p.25$. In this comparison our 'fiducial' model is the $2f_0$ run with a Prandtl
number of one. We observe sustained turbulence out to 150 orbits in all cases, implying that 
our effective Reynolds number (which should be boosted somewhat by our use of hyperviscosity 
rather than a regular viscosity) is high enough that sustained turbulence is possible even 
for ${\rm Pm_h} = 0.25$, i.e. that we are in the large Reynolds number regime described 
by \cite{fromang07b}. In fact, no clearly significant differences are seen between the 
${\rm Pm_h} = 0.25$ and ${\rm Pm_h} = 1$ runs. For ${\rm Pm_h} = 4$, on the other hand, 
we see an increase of about a factor of 2-2.5 in the magnetic field energy and angular 
momentum transport efficiency. Although the use of hyperviscosity and hyperresistivity 
means that our runs are not equivalent (and indeed are less physical) to those of 
\cite{lesur07} and \cite{fromang07b} this trend is consistent with those previous studies 
which show an increase in saturation level with increased Prandtl number. We borrow the 
same interpretation: values of ${\rm Pm > 1}$ result in strong viscous damping at scales 
larger than the magnetic field dissipation scale. This suppresses magnetic field 
dissipation at small scales, and favors an inverse cascade of magnetic energy to larger scales 
and an attendant increase in $\alpha$. 

While we see an increase in the saturation level for the higher Prandtl number simulation the 
the underlying behavior remains the same as that seen in the large scale forcing runs. For both
runs with altered Prandtl numbers the transport is still dominated by the Maxwell stresses and the
proportion of Maxwell to Reynolds stress remains the same. We find no significant differences in the
behavior of the field components or the power spectra except that they are amplified in the the higher
Prandtl number run.

\subsubsection{Large Scale Runs With Orbital Period Forcing} 
\begin{figure}
  \plotone{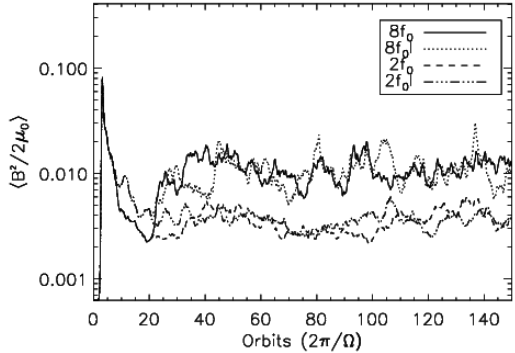}
  \caption{Evolution of the magnetic field energy density ($B^2/2\mu_0$) for the
simulations with orbital period forcing variations. The results are compared to those from the equivalent large
scale simulations in which the forcing was varied on the timescale of the simulation timestep. 
The results marked with an 'l' denote the orbital period simulations. No significant differences 
are seen.}
  \label{fig:magvolnf}
\end{figure}

In the large-scale forcing runs described above the vectors which describe the 
instantaneous turbulent driving are chosen anew at each timestep. The forcing is 
then temporally uncorrelated with any characteristic frequency of the disk. To 
check whether the results depend upon this feature we repeated some of the runs 
with the correlation time for the forcing set to be the orbital period\footnote{We 
adopted this timescale since it is a well-defined characteristic frequency 
for an accretion disk. Since it is also the resonant frequency, any effects 
due to changing the timescale are likely to be maximized by adopting orbital 
period scale forcing.}. We ran
simulations at two times and eight times the reference forcing amplitude with orbital period forcing
($2f_0l$ and $8f_0l$) and compared them to the standard runs with time step forcing variations. 
Figure~\ref{fig:magvolnf} shows the volume averaged magnetic energy density in the simulations with
orbital period forcing variation compared to their standard counterparts.  We find no discernible
differences in the magnetic field, transport coefficients, or spectra in the case of
orbital period variations. Both the behavior and the saturation amplitudes agree. Once again the
Maxwell stresses dominate the transport.

\subsection{Small Scale Forcing Runs}
We have also considered the effects of forcing on scales $\lambda \ll H$. Once 
again we begin by finding the strength of forcing that yields a saturated kinetic 
energy density that matches that obtained in a fiducial MRI-only simulation, 
except now we use a forcing function that injects power at wavenumbers around 
$k = 40$. The match we obtained was good at the $\approx 20$\% level. We found that 
the residual hydrodynamic $\alpha$ in the purely hydro run was substantially 
higher than in the case with large scale forcing, but still negligible for our 
purposes (being two order of magnitude below the MRI value).

\begin{figure}
  \plotone{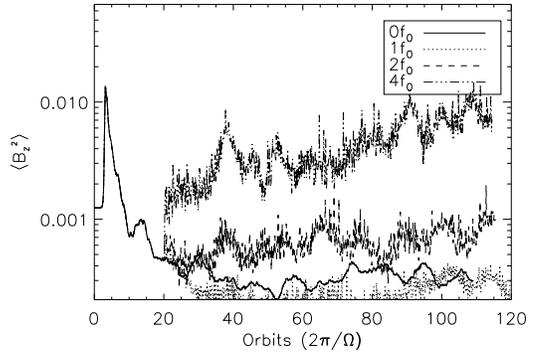} 
  \caption{Evolution of $B_z^2$ for
  the simulations with small scale forcing ($0sf_0,1sf_0,2sf_0, \& \, 4sf_0$).}
  \label{fig:bzbzs}
\end{figure}

\begin{figure}
  \plotone{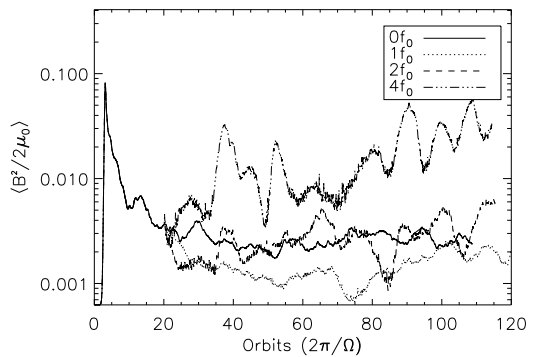}
\caption{Evolution of the magnetic field energy density $B^2/2\mu_0$ for the 
simulations with small scale forcing ($0sf_0,1sf_0,2sf_0, \& \, 4sf_0$).}
  \label{fig:bsqs}
\end{figure}

Figure~\ref{fig:bzbzs} and figure~\ref{fig:bsqs} show the time evolution of 
$B_z^2$ and $B^2/2 \mu_0$ for the small scale forcing simulations. We considered 
a more limited range of forcing strengths between $1 sf_0$ and $4 sf_0$. Within this 
range, there are some differences between the behavior of the disk subject to large and 
small scale forcing. Most notably, although relatively weak levels of forcing (the $2 sf_0$ run) 
do increase the value of the vertical magnetic field, this is not accompanied by any 
significant increase in the total magnetic field. The strongest level ($4 sf_0$) of forcing 
{\em does} amplify all components of the magnetic field. We also observe larger 
fluctuations in magnetic field strength as compared to the equivalent large-scale 
forcing simulations.

\begin{figure}
  \plotone{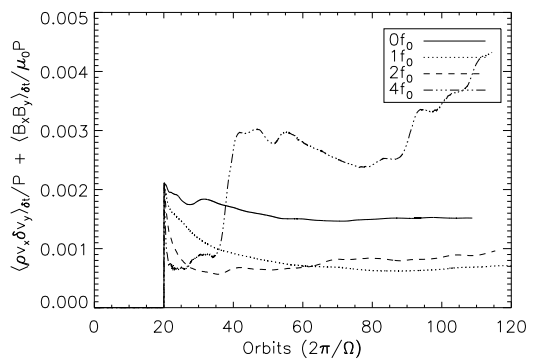}
  \caption{Running time average of $\alpha$ for the simulations with small scale forcing 
  ($0sf_0,1sf_0,2sf_0, \& \, 4sf_0$). The lowest two forcing strengths result in a 
  suppression of angular momentum transport.}
  \label{fig:alphatimes}
\end{figure}

The stress within the simulated patch of disk largely follows the total magnetic 
field strength. Figure~\ref{fig:alphatimes} shows the running time average of the 
equivalent $\alpha$ parameter for the different simulations. The average $\alpha$ is 
lower in the weakly forced runs ($1 sf_0$ and $2 sf_0$) than in the unforced 
simulation, and only shows significant enhancement in the $4 sf_0$ case. In this 
case it is clear that the value of $\alpha$ has not converged by the end of the 
simulation. We note that in all cases -- including those where the forcing 
appeared to suppress the MRI -- the transport remains dominated by Maxwell 
stresses.

\begin{figure}
  \plotone{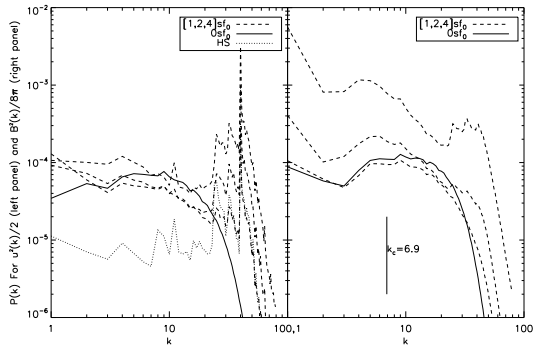}
  \caption{Power spectra of the kinetic energy (left panel) and the magnetic field energy density (right panel)
for simulations with small scale forcing. In these runs the forcing is applied at a wavenumber $k=40$.  
The spectra have been averaged over orbits 95-105. HS is the
hydrodynamic only simulation with small scale forcing, $0sf_0$ is the unforced MRI run, while the 
remaining curves correspond to the
small scale, forced MRI runs. Unlike in the large scale forcing runs Kolmogorov spectra are 
not observed at large scales, and there is a marked accumulation of power close to the 
injection wavenumber. Power in the large scale magnetic field is essentially unaltered by the 
forcing except in the highest amplitude forcing run.}
  \label{fig:ffts}
\end{figure}

Figure~\ref{fig:ffts} shows the power spectra of the kinetic energy (left panel) and 
magnetic energy density (right panel) for all of the small scale forcing simulations. 
The accumulation of power in the kinetic energy spectrum at scales close to the 
forcing scale of $k=40$ is very obvious in all the forced runs, and on this 
scale the power in fluid motions greatly exceeds that which would be generated 
by the MRI. In the hydro-only run power leaks out to larger scales, and results 
in a roughly flat power spectrum out to the box scale at $k=1$. The spectra 
of the magnetic field energy are rather different. In the two most weakly forced 
runs the forcing increases the power close to the forcing scale, but diminishes 
or leaves roughly unaltered the power on larger scales. In particular, for these 
runs there is no increase in the power on the scales -- comparable to the most unstable 
linear MRI modes -- that are presumably most important for field evolution in a disk. 
Only for the strongest level of forcing ($4 sf_0$) do we see a clear increase in the 
power across all of the scales accessible to the simulation.
   
Based on these results, we conjecture that the impact of hydrodynamic forcing on the 
MRI varies depending on the effect that the forcing has on magnetic fields with 
scales comparable to the most unstable MRI wavelengths in the disk. Large scale 
forcing efficiently enhances the vertical magnetic field in the disk at quite 
moderate amplitudes, and essentially always boosts the strength of angular momentum 
transport. Power from small scale forcing primarily affects the structure of the 
turbulence at scales close to and below the injection scale, and, at least if the 
forcing is weak, relatively little power cascades back to large scales. There is 
no enhancement of the field strength on the scales most important for growth of the 
MRI. Rather, the imposed turbulent motion on small scales acts to suppress the 
MRI, perhaps by enhancing magnetic field dissipation (the inverse of the Prandtl 
number effect discussed above), or perhaps by directly destroying the correlation 
between the radial and azimuthal field that is responsible for the Maxwell stress.
Only if the small scale forcing is rather strong does enough power reach larger 
scales, at which point the MRI can be enhanced via the same mechanisms as apply 
in the large scale forcing runs.

\section{Discussion}
In this paper we have used local shearing-box simulations of accretion 
disks to study the interaction between the magnetorotational instability and 
hydrodynamic turbulence in disks where both coexist. We have studied both 
large scale forcing, in which energy is injected at scales $\lambda \sim H$, 
the disk scale height, and to a more limited extent small scale forcing 
where $\lambda \ll H$. We find that the effect of the additional energy 
input from the hydrodynamic turbulence on the MRI varies depending upon 
both the strength of the forcing and, to some extent, on its spatial 
structure.

For large scale forcing the results are straightforward. When hydrodynamic 
forcing is ``weak" -- in the sense that the hydrodynamic forcing, on its 
own, yields a saturated kinetic energy density in the disk that is less 
than or equal to that generated by the MRI alone -- the MRI is essentially 
unaffected at the level of precision accessible to our simulations. This 
result is no surprise. The MRI is a robust instability, which is present 
within differentially rotating flows containing almost arbitrary 
magnetic field geometries \citep{balbus98}. Low amplitude random 
perturbations do not affect it. When stronger forcing is imposed, 
we find that both the saturation value of the magnetic field and the 
strength of angular momentum transport can be substantially boosted. 
The flow retains many of the characteristics of the MRI --- such as 
a similar ratio between magnetic field strength and $\alpha$, and a 
similar ratio of Maxwell to Reynolds stresses, even in a regime 
when the hydrodynamic forcing (which, by itself, yields no transport 
at all) is formally dominant. The application of heuristic dynamo 
arguments to the MRI is suspect \citep{balbus98}, but we tentatively 
attribute the numerical behavior to the more efficient regeneration of 
vertical field in the presence of hydrodynamic turbulence. 

The small scale forcing results are more nuanced, and given the limited 
number of simulations we have performed should be regarded as preliminary. 
In this regime we find that only the strongest level of forcing boosts the 
strength of angular momentum transport, whereas lower forcing levels 
actually suppress transport. We attribute this different behavior to the 
fact that there are two important considerations that affect the saturation 
amplitude of the MRI. One is the strength of the vertical magnetic field 
on relatively large scales (typically a fraction of $H$), similar to the 
most unstable linear MRI wavelengths. Enhancement of the vertical field 
on this scale -- which is readily accomplished with large scale 
forcing but which requires an inverse cascade in the small scale case -- 
boosts the strength of the MRI. The second is the dissipation scale. 
Turbulent driving that increases the amplitude of kinetic energy at 
or near this scale may enhance magnetic field dissipation, ultimately 
reducing the saturation level of magnetic fields in the disk as a 
whole. This is the inverse of the physical process invoked to explain 
the dependence of $\alpha$ on the magnetic Prandtl number \citep{lesur07,
fromang07b}. Small scale forcing may also directly destroy the correlations 
between $B_r$ and $B_\phi$ that result in Maxwell stress.

Our results allow us to infer which additional physical effects are 
likely to be able to affect the MRI in accretion disks. We first 
observe that if the additional turbulence is ``powered" by the MRI 
itself (i.e. if the turbulence derives energy from the gravitational 
potential well as a result of MRI-driven angular momentum transport), 
then generically it is unlikely to be as powerful as the MRI. Such 
turbulence will have at most a small effect on the magnitude and 
character of angular momentum transport. As an explicit example, we 
would not expect an MRI-active disk that was additionally unstable 
to convection to differ much from one in which the vertical structure  
was stable against convective motions.

We can also consider sources of turbulence that are independent 
of the MRI, in the sense that they would exist even in a (hypothetical) 
disk that was absent magnetic fields entirely. Physical examples 
include self-gravity, thermal instability, and turbulence stirred 
up by the interaction between the gaseous and solid components of 
protoplanetary disks. There is no reason why these sources of 
turbulence should not generate fluid motions of greater amplitude 
than those produced by the MRI (this is likely to be the case in 
self-gravitating disks near the fragmentation threshold, and 
locally in some regions of protoplanetary disks). In this regime, 
our results suggest that the MRI will have a non-trivial  
interaction with the hydrodynamic turbulence. If the forcing 
occurs at scales comparable to $H$ or larger, we find that the 
interaction will likely boost the strength of angular momentum 
transport. Even for quite violent forcing -- up to an order of 
magnitude in excess of that required to produce parity 
with the MRI -- we find that it is 
most accurate to think of the coupled system as one with 
boosted MHD turbulence, rather than as a hydrodynamic system 
passively advecting magnetic fields. Conversely, small scale 
forcing, unless it is very strong, may actually suppress the 
saturation level of magnetic fields in the disk and their 
associated angular momentum transport.

\acknowledgements

This work was supported by NASA under grants NNG04GL01G and NNX07AH08G 
from the Astrophysics Theory Program, and 
by the NSF under grant AST~0407040. Simulations were run on the NSF's 
TeraGrid platforms, and on NASA's Columbia system. 
We thank Brian Morsony for useful discussions, and the referee for 
many helpful suggestions including the idea of looking at small 
scale driving. We are grateful to Peter Ruprecht, James McKown, and the support 
teams at NCSA and Columbia for invaluable assistance with the computations, and the 
authors and users of {\sc PENCIL} for help with the code.











\clearpage










\end{document}